\documentclass[11pt,twoside]{article}

\usepackage{asp2014}

\aspSuppressVolSlug
\resetcounters

\begin{document}

\title{Processing All-Sky Images At Scale On The Amazon Cloud: A HiPS Example}


\author{G. Bruce~Berriman$^1$, John C.~Good$^2$}
\affil{$^1$, Caltech/IPAC-NExScI, Pasadena, CA 91125, USA; \email{gbb@ipac.caltech.edu}}
\affil{$^2$ Caltech/IPAC-NExScI, Pasadena, CA 91125, USA}
\paperauthor{G.~Bruce Berriman}{gbb@ipac.caltech.edu}{0000-0001-8388-534X}{Caltech}{IPAC/NExScI}{Pasadena}{CA}{91125}{USA}
\paperauthor{J.~C.~Good}{jcg@ipac.caltech.edu}{ }{Caltech}{IPAC/NExScI}{Pasadena}{CA}{91125}{USA}



\begin{abstract}
We report here on a project that has developed a practical approach to processing all-sky image collections on cloud platforms, using as an exemplar application the creation of three-color Hierarchical Progressive Survey (HiPS) maps of the 2MASS data set with the Montage Image Mosaic Engine on Amazon Web Services. We will emphasize issues that must be considered by scientists wishing to use cloud platforms to perform such parallel processing, so providing a guide for scientists wishing to exploit cloud platforms for similar large-scale processing. A HiPS map is based on the HEALPix sky-tiling scheme. Progressive zooming of a HiPS map reveals an image sampled at ever smaller or larger spatial scales that are defined by the HEALPix standard. Briefly, the approach used by Montage involves creating a base mosaic at the lowest required HEALPix level, usually chosen to match as closely as possible the spatial sampling of the input images, then cutting out the HiPS cells in PNG format from this mosaic. The process is repeated at successive HEALPix levels to create a nested collection of FITS files, from which PNG files are created that are shown in HiPS viewers. Stretching FITS files to produce PNGs is based on an image histogram. For composite regions (up and including the whole sky), the histograms for each tile can be combined to create a composite histogram for the region. Using this single histogram for each of the individual FITS files means all the PNGs are on the same brightness scale and displaying them side by side in a HiPS viewer produces a continuous uniform map across the entire sky.

All of the processing just described can be readily performed in parallel on AWS instances. To create the HiPS maps on AWS, jobs were set up with a Docker container that contains the requisite data software components, including modules added to streamline processing on cloud platforms---including adjusting for inter-image background variations and developing a global model for visualization stretches. Jobs are set up and run with the Amazon Web Services (AWS) Batch processing mode, which spins up server instances as needed, pulling from a pool of predefined job scripts. When a job is done, it either another job from the pool or shuts the instance down. This approach minimizes having idle instances, which would still incur charges even when not processing. A set of script generators developed for this project create, by design, simple scripts that are handed to the instances to run jobs inside the containers. Processing the whole sky at three wavelengths requires about ten thousand such jobs. We will discuss processing times and costs.

\end{abstract}



\section{Introduction}

This paper describes the creation of three-color Hierarchical Progressive Survey (HiPS) maps of the all-sky 2MASS images on the Amazon Elastic Cloud 2 of Amazon Web Services (AWS), with the Montage Image Mosaic Engine \citep{2017PASP..129e8006B}. While not a large collection by current standards, the 4,879,128 calibrated FITS images delivered by 2MASS (8 TB volume) at J (1.25 $\micron$), (1.65 $\micron$), and (2.2 $\micron$) are large enough to demonstrate the process at a reasonable cost. The methodology described here easily scales up to much larger image collections and is therefore an exemplar of how HiPS maps can be created at scale. 

HiPS maps offer a way of visualizing the sky at progressively higher spatial sampling by zooming inwards in a HiPS-compliant viewer, such as Aladin Lite \citep{2015A&A...578A.114F}. HiPS has been recommended as a standard by the International Virtual Observatory Alliance (IVOA) \footnote{\url{https://www.ivoa.net/documents/HiPS/}}. Production of HiPS maps is straightforward but often computationally expensive. A mosaic of the sky is stored as a hierarchy of FITS files. The hierarchy is based on the Hierarchical Equal Area isoLatitude Pixelation of a Sphere (HEALPix), a subdivision of a spherical surface in which each pixel has the same surface area \citep{2005ApJ...622..759G}. The mosaic is partitioned into a tree of PNG files that are used for visualization in HiPS-compliant viewers. 

\section{HiPS Maps at Infrared Wavelengths}
The methodology developed at CDS for computing HiPS maps, Hipsgen, emphasizes efficiency and simplicity: it builds a collection of files for 0.2-arcsecond pixels for the entire sky in roughly two days. As of this writing, this has led to the delivery of an impressive 1246 maps from 20 HiPs servers worldwide. The disadvantage of Hipsgen arises in processing infrared data sets; the constant background it uses is often insufficient to handle much higher and more variable backgrounds than seen in the optical. In the  2MASS images, OH-emission varies rapidly on timescales of 10 minutes and is particularly prominent in the H-band. Montage models this background radiation across images \citep{2017PASP..129e8006B}, but at greater computational expense than in Hipsgen. Moreover, Montage creates mosaics, supports HEALPix, supports an adaptive stretch for image display, and is highly parallelizable. Thus, Montage is a good choice for creating HiPS maps at scale at infrared wavelengths \citep{2017PASP..129e8006B}; \citep{2020ASPC..527..235B}.

\section{Building HiPS Maps on the Amazon Cloud}

The processing was performed with AWS Batch, which provisions servers as needed, starts a job from a pool of predefined job scripts, shuts down or starts another job as needed, and cleans up after itself. Instances are live only when operating, and this automatically manages processing costs. The Montage software was built in a Docker container, the preferred mechanism for AWS batch processing.

The process of creating the HiPS tiles was all done in parallel. The sky was divided into 3328 tiles, at 1-arcsecond resolution. New modules were required to perform processing on the cloud; these are summarized in Table 1. In particular, a set of script generators were developed to create scripts to run jobs, including creating the mosaic, shrinking the original images by successive factors of x2, and then building PNGs; the last item took advantage of the adaptive histogram-based image stretch used in Montage.

\begin{table}
    \centering
    \begin{tabular}{ll}
        \bf Module & \bf Description \\
         & \\
         mHPXHdr   &  Create HiPS-compliant header \\
         mHPXMosaicScripts & Script Builder  \\
         mHPXShrinkScripts   & Create scripts for making lower-resolution images \\
         mHIPSTiles & Creates 512 x 512 cutouts \\
         mHIPSTileScripts  & Scripts to create 512 x 512 cutouts with mHIPSTiles \\
         mHIPSPNG & Converts FITS files to PNG files \\
         mHiPSPNGScripts  & Scripts to convert FITS to PNGs with mHiPSPNG \\
         mHiPSAllSky & Creates a composite image of low-resolution tiles  \\
    \end{tabular}
    \caption{New Montage modules to support the creation of HIPS maps}
    \label{tab:my_label}
\end{table}
    
Table 2 summarizes the costs of creating and downloading the PNG files. The biggest cost is for processing, which incurs roughly three-quarters of the total costs of \$1,651. Note, though, that downloading the FITS files---which are not required by HiPS viewers---will incur an additional cost of \$2,700. There is essentially no cost in storing the input files because they are deleted immediately after they are processed.

\begin{table}
    \centering
    \begin{tabular}{llllr}
      \bf Resource   & \bf Base Cost & \bf Time  & \bf Volume  & \bf Cost \\
         &  &  &  & \\
      Processing    & \$0.0416/hour  & 9984 hours & ... & \$1,216 \\
      S3 Storage   & \$0.23 GB/month &2 weeks  & 30 TB & \$345  \\
       Download PNGs   & \$0.09/GB  & ... & 1 TB  &\$90  \\
        \bf Grand Total &  &  &  &\bf \$1,651   \\
    \end{tabular}
    \caption{Summary of the costs of creating the 2MASS three-color HiPS maps. Downloading the 30 TB of FITS files would incur an additional cost of \$2,700}
    \label{tab:my_label}
\end{table}

Table 2 refers only to the cost of processing itself. There can be additional costs of up to 20-30 \% more in learning how to use the AWS Batch system and operating it:

\begin{itemize}
    \item By default, AWS only allows a small number of instances per session; AWS must grant permission for access to more instances.
    \item Users determine empirically what instance type, memory allocation, and storage are needed for AWS Batch, usually by trial and error. AWS Batch itself determines the exact instance it uses.
    \item The default algorithm used by AWS Batch has little transparency. In the case of 2MASS, we ran out of space even when a storage limit was set at a high value: AWS Batch may have high storage overheads associated with it. 
    \item The CloudWatch alert system we used to monitor jobs appears tailored to system administrators rather than users. 
    \item Bad jobs needed to be shut down manually via the AWS Command Line Interface.
    \item A small difference in the C compiler between Linux and our Mac development machines meant we had to test Montage on AWS.
    
\end{itemize}

Figure 1 shows a three-color image of M17, part of the HiPS map computed on AWS. (Residual background effects are present because the rapid background variations are not fully rectified by Montage.)

\articlefigure{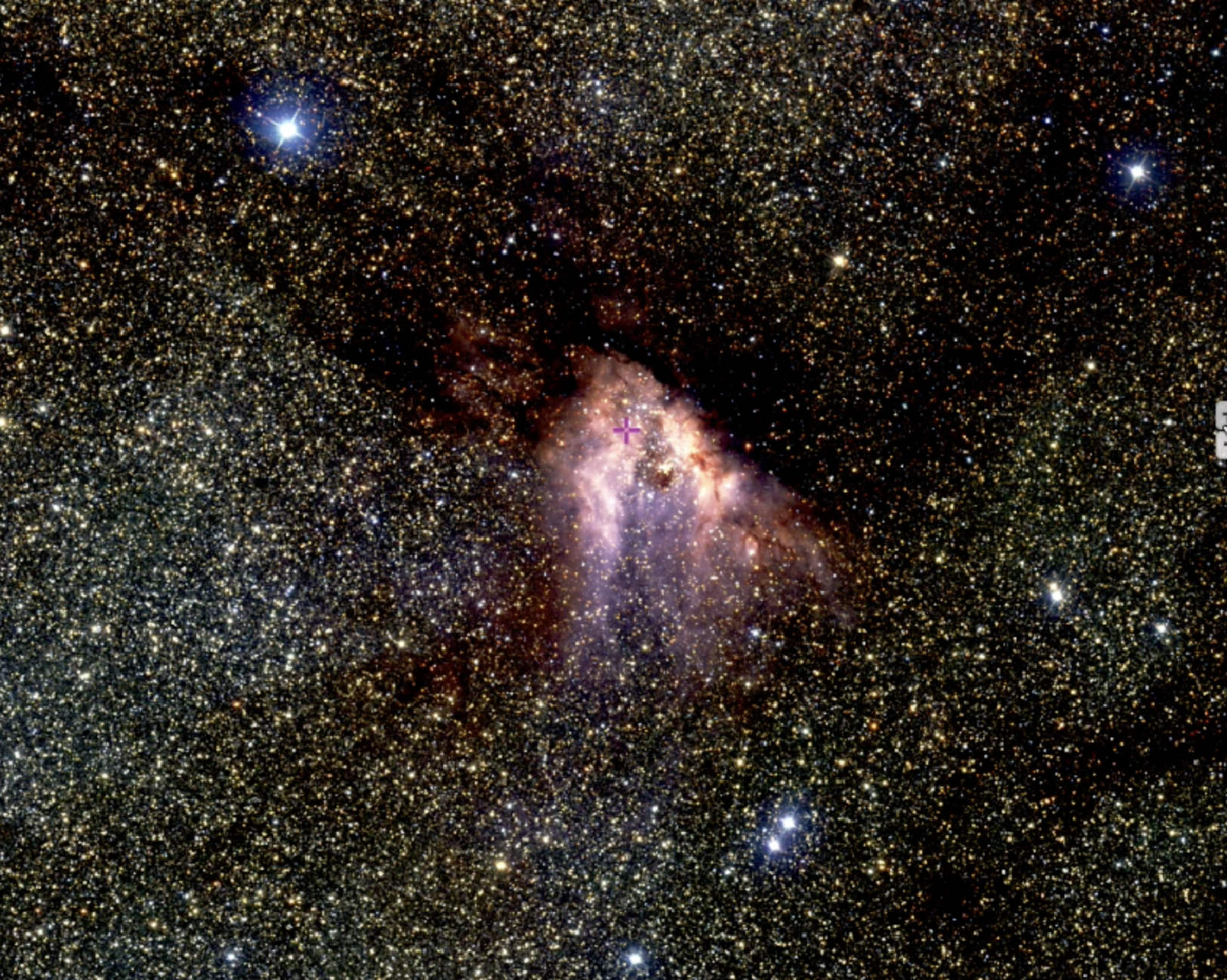}{ex_fig1}{Three-color 2MASS HiPS image of M17 shown at HEALPix level 10 (3.4 arcmin)  and visualized through  Aladin Lite}

\acknowledgments We thank the Caltech IST/AWS AI4Science Cloud Credit Program for a generous allocation of cloud credits. This research made use of Montage. It is funded by the National Science Foundation under Grant Numbers ACI-1440620, 1642453, 1835379.

\bibliography{C808}  


\end{document}